\documentclass[aps,prl,preprint,floatfix,showpacs,superscriptaddress]{revtex4}

\usepackage{amsmath}
\usepackage{graphicx}
\usepackage{dcolumn}
\usepackage{bm}

\newcommand{\tbh}[3]{\multicolumn{#1}{#2}{\textbf{#3}}}
\newcommand{\tbhc}[1]{\multicolumn{1}{c}{\textbf{#1}}}

\begin{document}

\title{\textit{Ab initio} calculations of inelastic losses and optical
constants}

\date{\today}

\author{J. J. Rehr}
\affiliation{Department of Physics, University of Washington, Seattle, WA 98195}

\author{J. J. Kas}

\affiliation{Department of Physics, University of Washington, Seattle, WA 98195}

\author{M. P. Prange}
\affiliation{Department of Physics, University of Washington, Seattle, WA 98195}

\author{F. D. Vila}
\affiliation{Department of Physics, University of Washington, Seattle, WA 98195}

\author{A. L. Ankudinov}
\affiliation{Department of Physics, University of Washington, Seattle, WA 98195}

\author{L. W. Campbell}
\affiliation{Department of Physics, State University of New York, Geneseo, NY
14454}

\author{A. P. Sorini}
\affiliation{Department of Physics, University of Washington, Seattle, WA 98195}

\begin{abstract}
{\it Ab initio} approaches are introduced for calculations of inelastic losses
and vibrational damping in core level x-ray and electron spectroscopies. From
the dielectric response function we obtain system-dependent self-energies,
inelastic mean free paths,  and losses due to multiple-electron excitations,
while from the dynamical matrix we obtain phonon spectra and Debye-Waller
factors. These developments yield various spectra and optical constants from the
UV to x-ray energies in aperiodic materials, and significantly improve both the
near edge and extended fine structure.
\end{abstract}

\maketitle

Theories of x-ray and electron spectra and other ``optical constants"
have become increasingly
sophisticated \cite{onida02,rehr00}. For example, treatments of such spectra
can  now go beyond the independent-particle approximation, e.g.,  with
time-dependent density functional theory (TDDFT) or the Bethe-Salpeter equation
(BSE) \cite{shirley01,ankudinov05}. However, less attention has been devoted to
the effects of inelastic losses and vibrational damping, especially at high
energies. These losses are important both in spectroscopy and other applications
\cite{powell99,dingfelder05}. Thus current treatments
\cite{horsch88,schattke02,rehr00}  often utilize simplified models, e.g.,
semi-empirical or electron-gas self-energies and Einstein or Debye models for
vibrations, with considerable variations in the results.

To address these difficulties we introduce in this Letter {\it ab initio}
approaches for these many-body damping effects. We first present an approach for
calculating the dielectric response and optical properties within a real-space
Green's function (RSGF) formalism \cite{rehr00,ankudinov05}. From these results,
we derive a many-pole model for the dielectric function, which is then used to
obtain  photoelectron self-energies, inelastic mean free paths (IMFPs), and
contributions from multi-electron excitations.  Next, from calculations of the
dynamical matrix we obtain phonon spectra and quantitative Debye-Waller (DW)
factors.  These developments enable a number of applications, e.g., improved
calculations of spectra and inelastic losses from the UV
to hard-x-ray limits.  

A key physical quantity in these calculations is the dielectric response
function $\chi(\omega)$ \cite{onida02}. In the long wave-length limit, various
spectra  are related to $\chi(\omega)$ through the local atomic polarizability
$\alpha(\omega)$ \cite{zangwill80},
\begin{equation}
\label{eq:alpha}
\begin{split}
  \alpha (\omega) &\equiv \int d{\bf r} d{\bf r'}\, d^{\dagger}({\bf r})\,
   \chi({\bf r,r'},\omega)\, d({\bf r}) \\ 
                  &=      \sum_{i,L,L'} \tilde M_{i,L}\,\langle G_{L,L'}(E)
                  \rangle\, \tilde M_{L'i} \, .
\end{split}
\end{equation}
Here, $\tilde M_{i,L}$ are screened dipole matrix elements between occupied core
states $|i\rangle$ and  final scattering states $|L,0\rangle$, and $G_{L,L'}(E)$
are matrix elements of the photoelectron Green's function in an angular momentum
and site basis \cite{ankudinov05}. The operator $d({\bf r})$ represents the
dipole coupling to the photon field, and the brackets $\langle \cdots \rangle$
denote a configurational average, which can be expressed in terms of DW factors.
\begin{figure}
\includegraphics[scale=0.35,clip]{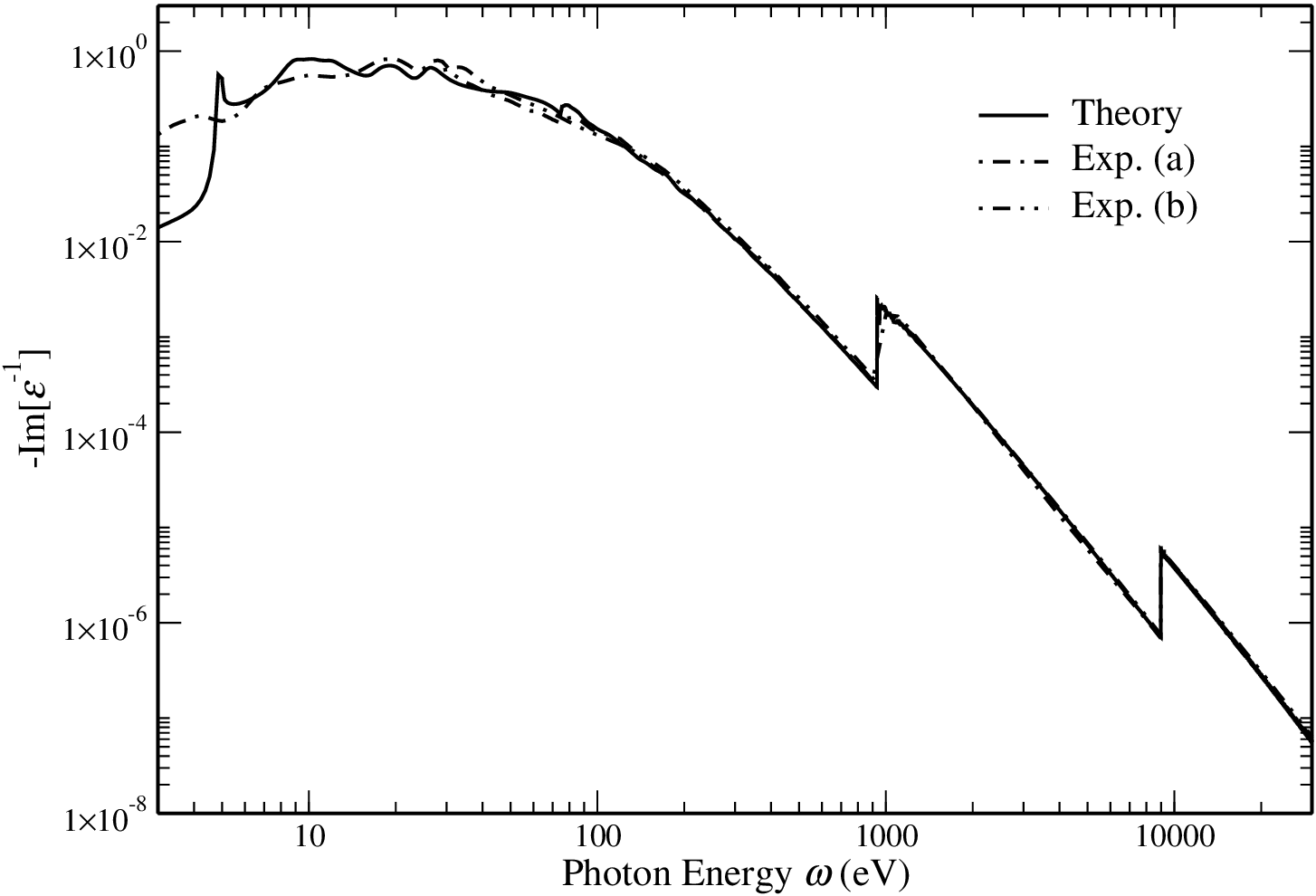}
\caption{\label{fig:epsm1cu}
Energy loss function $-{\rm Im}\, [\epsilon^{-1}]$ for fcc Cu \textit{vs}
photon energy $\omega$ from this work (solid), and from experiment
(a) (dash-dot) \protect\cite{hagemann75} and
(b) (dash-dot-dot) \protect\cite{henke93}.
}
\end{figure}
From $\alpha(\omega)$ we obtain the dielectric constant $\epsilon(\omega) =
1+4\pi n\alpha(\omega)$, where $n=N/V$ is the atomic number density
\cite{prangeetal06}. Other optical constants (e.g., absorption coefficient
$\mu$, reflectivity $\cal R$, anomalous x-ray scattering amplitudes $f_1+if_2$,
etc.) are related to $\epsilon\equiv\epsilon_1 + i \epsilon_2$. Typical results,
e.g., for the energy loss function $- {\rm
Im}\,[\epsilon^{-1}]=\epsilon_2/[\epsilon_1^2+\epsilon_2^2]$ for Cu
(Fig.~\ref{fig:epsm1cu}), are in good agreement with experiment
\cite{hagemann75,henke93} in the UV and beyond. This and other examples
based on our approach are tabulated on the WWW \cite{prangetables05}.

The quantities in Eq.~(\ref{eq:alpha}) are calculated using an extension of our
RSGF code \cite{prangeetal06} that includes corrections to the
independent-particle approximation \cite{ankudinov05}. Compared to other
approaches, the RSGF method is advantageous for core-level spectra since it
avoids an explicit calculation of energy eigenstates and is applicable to
periodic and aperiodic systems alike. For each core state $|i\rangle$, we first
calculate the atomic-like background with no scattering; fine structure is then
added with full multiple scattering (MS) to about 50 eV, and then with the MS
path expansion to about 1500 eV above each edge. In our code both
$G(E)=1/(E-h'+i0^+)$ and the scattering states $|L,0\rangle=R_l(r) Y_L({\bf \hat
r})$ are obtained with a self-consistent final-state, muffin-tin Hamiltonian
$h'=p^2/2 + V'_{coul} + \Sigma(E)$. This Hamiltonian includes a screened
core-hole, an energy-dependent self energy $\Sigma(E)$, and the core-hole
lifetime $\Gamma$ (we use Hartree atomic units $e=\hbar=m=1$).  Our screened
core-hole agrees well with that from the random phase approximation, and gives
results for core-level spectra which are a good approximation to the BSE
\cite{rehr05}. The screened dipole operators in $\tilde M$ account for the
induced local dipole fields \cite{zangwill80}. For metals, the low-frequency
behavior of $\epsilon(\omega)$ is approximated by an additive Drude term;
otherwise there are no adjustable parameters.

\begin{figure}
\includegraphics[scale=0.35,clip]{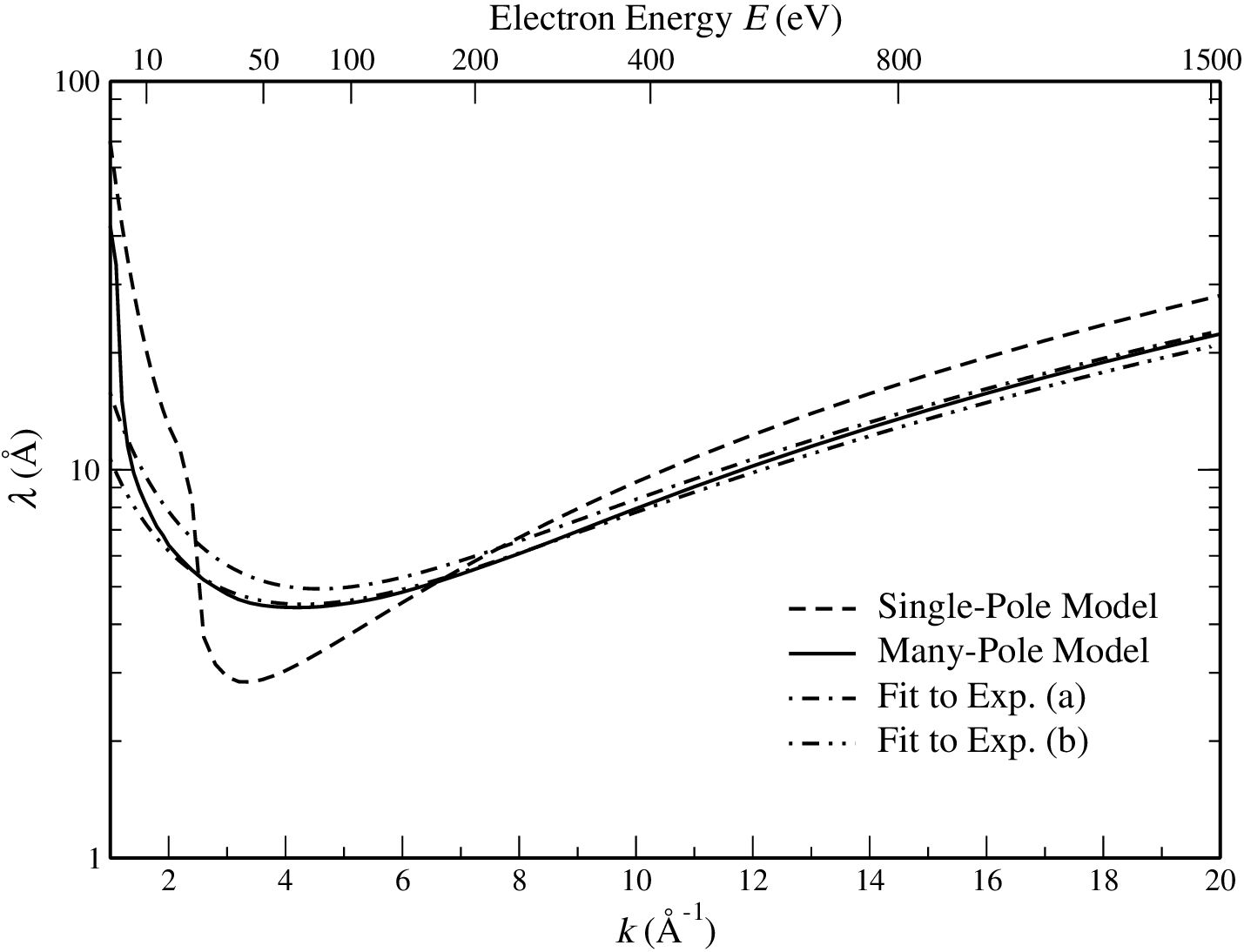}
\caption{\label{fig:imfpcu}
Inelastic mean free path $\lambda$ for fcc Cu \textit{vs} electron wave number
$k=(2E)^{1/2}$ and kinetic energy $E$  for our many-pole model (solid), a single
plasmon-pole model (dashes); and fits to experiment (dash-dots) (a)
\protect\cite{kwei93} and (b)  \protect\cite{powell99}.
}
\end{figure}

From the dielectric response, we obtain the self-energy $\Sigma(E)$ using the
``GW approximation" \cite{hedin69},
\begin{equation}
\label{eq:gw_app}
    \Sigma(E)=i\int { \frac{d\omega}{2\pi} } 
                G(E-\omega)W(\omega)e^{-i\delta\omega},
\end{equation}
where $W=\epsilon^{-1}(\omega)V$ is the screened coulomb interaction, and matrix
indices (${\bf r,r'}$) are suppressed. For $G(E)$, we use the free Green's
function, ignoring MS terms. Hence our $\Sigma(E)$ represents an {\it average}
self-energy, which is nevertheless adequate for many applications. Current GW
implementations for core spectra \cite{rehr00} typically use a single-plasmon
pole approximation for the dielectric function. Although computationally
efficient, Fig.~1 shows that a single pole may not be a good representation.
Thus, to obtain improved self-energies that exploit the efficiency of pole
models, we match our {\it ab initio} $\epsilon(\omega)$ to a
many-pole model, with poles spread over a broad spectral range, 
\begin{equation}
\label{eq:imepsm1mp}
  - \mathrm{Im}\left[\epsilon^{-1}({\bf{q}},\omega)\right]=
  \frac{\pi}{2}\sum_{j}{g_j \omega_{j}
   ~\delta \left(\omega - \omega_{j}\left({\bf {q}}\right)\right)}.
\end{equation}
Here $g_j = \left(2 \Delta \omega_j/{\pi \omega_j}\right) ~\mathrm{Im} \left[
\epsilon^{-1}\left(\omega_j\right)\right]$ is the strength of pole $j$ and
$\Delta\omega_j$ is the pole spacing. For simplicity, the momentum dependence of
each pole is approximated by a polynomial in $q^2$: $\omega_j({\bf
q})=[\omega_j^2+q^2/3 + q^4/4]^{1/2}$ \cite{lundqvist67}. This approximation is
roughly consistent with explicit calculations \cite{soininen05}, but the precise
dispersion is not critical since ${\bf q}$ and $\omega$ are integrated over
\cite{lundqvist67}. Many-pole models \cite{soininen03} and other methods
\cite{rubio99} have been developed for more accurate self-energy calculations;
however, these methods are computationally demanding and not applicable at high
photoelectron energies (e.g., for the extended x-ray absorption fine structure
(EXAFS)). Our model is similar to that of Ref.~\cite{horsch88}, but does not
rely on empirical optical constants.  Thus using Eq.~(\ref{eq:imepsm1mp}) in
Eq.~(\ref{eq:gw_app}), we obtain an efficient approximation for  the self-energy
as a sum over single-pole models with excitation energies $\omega_{j}$, i.e.,
$\Sigma (E) = \sum_{j}{g_j~\Sigma(E,\omega_j)}$. As an application we have
calculated the IMFP $\lambda = (E/2)^{1/2}/|{\rm Im}\,\Sigma(E)|$  for various
materials \cite{prangetables05}. Fig.\ \ref{fig:imfpcu} shows $\lambda$ for fcc
Cu using a 100-pole model  matched to the loss function in
Fig.~\ref{fig:epsm1cu}. Clearly this model corrects the excessive loss of the
single-pole model below 100 eV, and yields quantitative agreement with
semi-empirical fits  to experiment \cite{kwei93,powell99}. Our self-energy is
also in reasonable agreement with that of Ref.~\cite{soininen03}. A similar
approach \cite{kasetal06} can be applied to inelastic electron scattering
\cite{dingfelder05} and to photoemission spectra \cite{schattke02}.

\begin{figure}
\includegraphics[scale=0.35,clip]{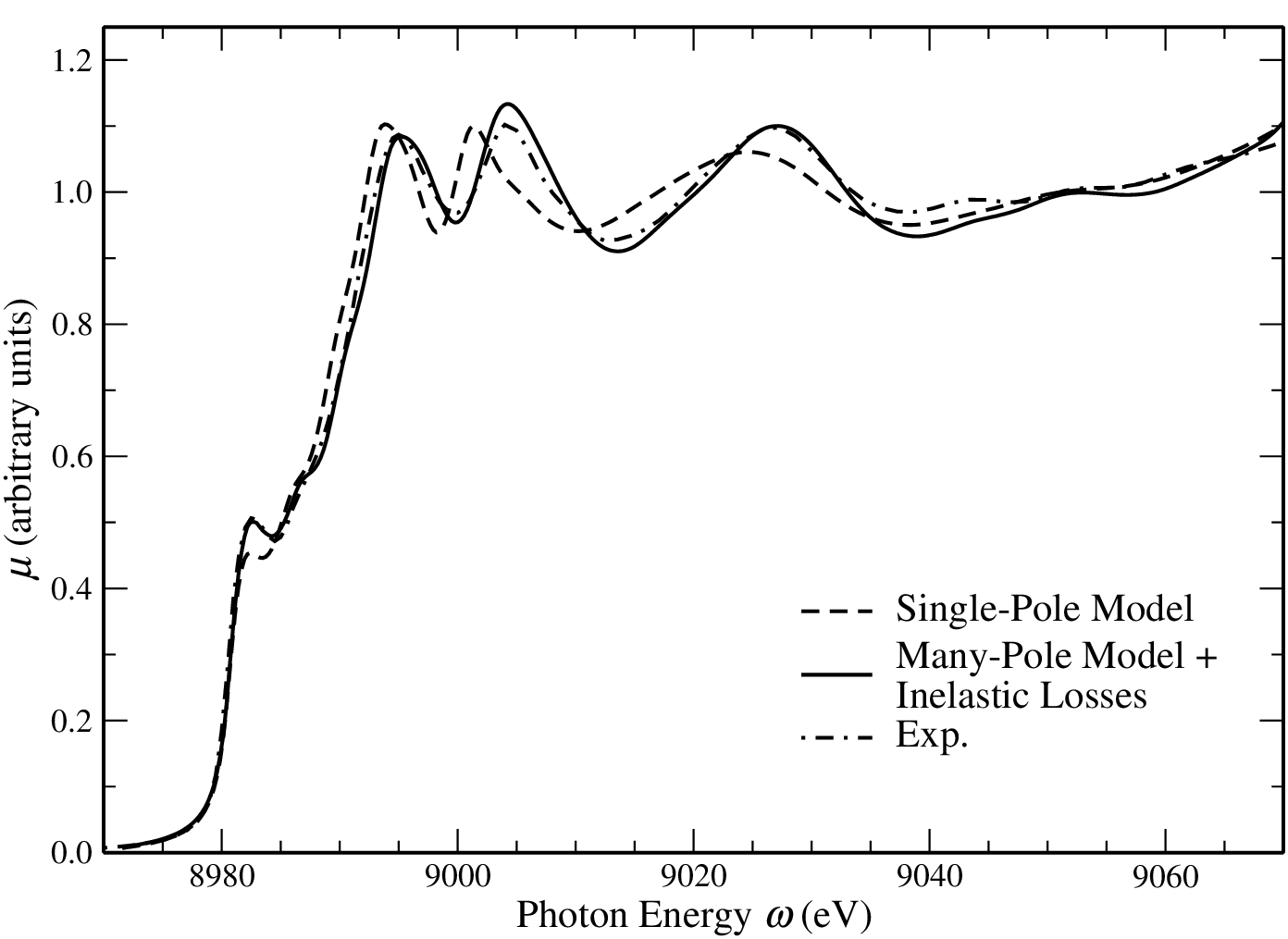}
\caption{\label{fig:xmu}
X-ray absorption $\mu$ {\it vs} photon energy $\omega$ for fcc Cu for our
many-pole model with inelastic losses (solid); a single-pole model (dashes); and
experiment (dash-dots) \protect\cite{newville94}.
}
\end{figure}

\begin{figure}
\includegraphics[scale=0.35,clip]{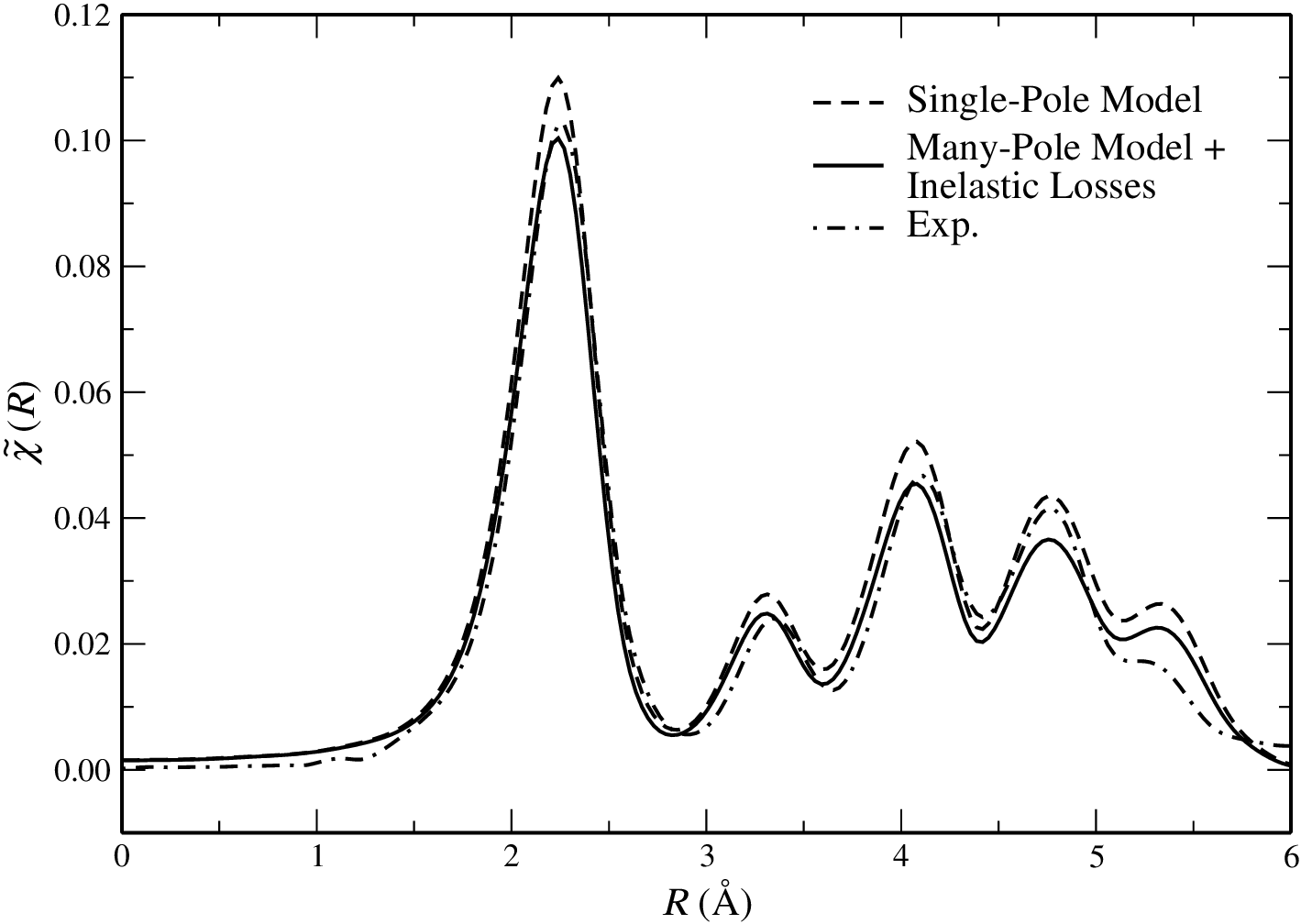}
\caption{\label{fig:inlosscu}
Fourier transform of the Cu K-edge EXAFS $\tilde\chi(R)$  at 10K vs half MS
path-length $R$ with (solid), and without inelastic losses (dashes), and from
experiment (dash-dots) \protect\cite{newville94}.
}
\end{figure}

Our many-pole model also permits calculations of inelastic losses
due to multi-electron (e.g.\  ``shake-up" and ``shake-off") excitations. These
excitations correspond to satellites beyond the quasi-particle peak
in the {\it spectral function} $A=(-1/\pi)\,{\rm Im}\, G_{\rm eff}$.
 Here $G_{\rm eff}$ is an effective,
one-particle propagator, which can be calculated using a generalization of the
GW approximation with similar ingredients \cite{campbell02}. In addition to
intrinsic losses, $G_{\rm eff}$ contains energy-dependent interference terms
which tend to suppress the satellites. As shown in Ref.~{\cite{campbell02}},
these excitations can be included in $\mu(\omega)$ in terms of a convolution
of a normalized
spectral function $\tilde A$ and the quasi-particle absorption coefficient
$\mu_{qp}$ in the absence of satellites,
\begin{equation}
\label{eq:mumb}
  \mu(\omega)=\int_0^{\infty} \!d\omega'
\tilde{A} (\omega,\omega')\mu_{qp}(\omega-\omega') \equiv
\langle \mu_{qp}(\omega)\rangle,
\end{equation}
where $\omega'$ is the excitation energy. Likewise, the net EXAFS is given by a
convolution of $\tilde A$ with the quasi-particle fine structure. Thus for each
path $R$ in the MS path expansion, the convolution over the oscillatory fine
structure yields an amplitude reduction factor $S_R^2(\omega)$ and (since $A$ is
asymmetric) a negative phase shift $\delta_R(\omega)$,
\begin{equation}
\label{chipath}
 \langle e^{2ikR} \rangle  = S_R^2(\omega)\, e^{2ikR + \delta_R(\omega)}.
\end{equation}
For the first shell of Cu, we find that $S_R^2(\omega)$ and $\delta_R(\omega)$
cross-over smoothly from the adiabatic (or quasi-particle) limit at threshold to
nearly constant values ($\sim$0.92 and $\sim - 0.2$ rad, respectively) in the
EXAFS, consistent with experiment $S_R^2\approx 0.9$ \cite{campbell02}. 
Clearly, our treatment of losses yields significant improvements in
amplitude and phase compared to the single-pole model, both for the near edge
spectra (XANES) (Fig.~\ref{fig:xmu}) and the EXAFS (Fig.~\ref{fig:inlosscu}).
Additional details are given elsewhere \cite{kasetal06}.

Finally, we present an {\it ab initio} approach for calculating the effects of
vibrational damping on the spectra. Our method is based on calculations of the
dynamical matrix ${\bf D}$ together with a Lanczos algorithm  for the phonon
spectra \cite{poiarkova01,krappe02}.  Instead of relying on spring-constants or
semi-empirical approximations, we use first principles electronic structure
methods to determine the real space matrix elements of ${\bf D}$,
\begin{equation}
\label{eq:rs_dm}
  D_{jl\alpha,j'l'\beta} =
  {\left( M_{j} M_{j'} \right)^{-1/2}} ~
  \frac{\partial^{2} U}{\partial u_{jl\alpha}\partial u_{j'l'\beta}},
\end{equation}
where $U$ is the total internal energy of the system and $u_{jl\alpha}$ is the
$\alpha = \left\{ x,y,z \right\}$ Cartesian displacement of atom $j$ of mass
$M_j$ in unit cell $l$. The DW factor $\exp\left(-2\sigma^2_R k^2\right)$ for a
given MS path $R$ is related to the the second cumulant $\sigma^2_R\equiv\langle
(\delta R)^2\rangle$ of the vibrational distribution function for that path
\cite{poiarkova01},
\begin{equation}
\label{eq:dw_fac}
  \sigma^2_R(T) = \frac{\hbar}{\mu_R}
                  \int_{0}^{\infty}{\rho_R\left(\omega^2\right)
  \mathrm{coth} \frac{\beta\hbar\omega}{2}}  d\omega .
\end{equation}
Here $\mu_R$ is a reduced mass for path $R$, $\beta = 1/{k_B T}$ and
$\rho_R\left(\omega^2\right) \equiv \left< Q_R\left|\delta\left(\omega^2
\bf{1}-{\bf D} \right)\right|Q_R\right>$ is the projected density of phonon
modes, which is approximated by a six-step Lanczos recursion, starting from a
path displacement seed state $|Q_R\rangle$. Other phonon spectra can be obtained
by varying the seed \cite{vila06}. Although our approach is applicable both to
molecules and condensed systems, we present results here only for two
crystalline examples. For these systems the $D_{jl\alpha,j'l'\beta}$ are
obtained using ABINIT \cite{gonze02}. Briefly, the reciprocal-space dynamical
matrix $\tilde{D}_{j\alpha,j'\beta} \left( \mathbf q \right)$ \cite{gonze97}  is
calculated on a $4\times4\times4$ grid of ${\bf q}$-points and smoothly
interpolated within the Brillouin zone. The real-space force constants (i.e.,
second derivatives of $U$) are then obtained from an inverse Fourier transform.
These calculations use Troullier-Martins-FHI pseudopotentials \cite{fuchs99}
with energy cutoffs of 60 and 15 Hartrees for Cu and Ge respectively,  while an
$8\times8\times8$ Monkhorst-Pack grid is used for the electronic energies. Table
\ref{tab:DW_fact} presents results obtained with local density approximation
(LDA)
\begin{table}
\caption{
Debye-Waller factors (in $10^{-3}$ \AA$^2$) for fcc Cu and dia Ge at 295K for
leading near-neighbor shells.
}
\label{tab:DW_fact}
\begin{ruledtabular}
\begin{tabular} {cdddd}
             & \tbh{3}{c}{Cu}                                     \\
               \cline{2-5}                                        \\[-8pt]
\tbhc{Shell} & \tbhc{LDA} & \tbhc{GGA} & \tbhc{CD}  & \tbhc{Exp.
\textnormal{\cite{stern80}}} \\
     1       &    8.55    &    7.15    &    8.91    &	    7.93\pm0.16\\
     2       &   11.06    &    9.50    &   10.99    &	   11.08\pm0.50\\
     3       &   10.80    &    9.14    &   11.32    &	    9.58\pm0.60\\[4pt]
& \tbh{3}{c}{Ge}                                                       \\
               \cline{2-5}                                             \\[-8pt]
\tbhc{Shell} & \tbhc{LDA} & \tbhc{GGA} & \tbhc{CD}  & \tbhc{Exp.
\textnormal{\cite{stern80}}}                                           \\
     1       &   3.77	  &    3.41    &   5.12     &	    3.5\pm0.15 \\
     2       &  10.21	  &   11.59    &   7.43     &	    9.3\pm1.1  \\[4pt]
\end{tabular}
\end{ruledtabular}
\end{table}
\cite{perdew92} and generalized gradient approximation (GGA) \cite{perdew96}
functionals.  The results all lie within about $\pm 10\% $ of experiment
\cite{stern80},  with the LDA usually closer. For comparison we show results for
the correlated Debye (CD) model \cite{rehr00}, which works well for Cu, but
gives significant errors for anisotropic structures like Ge. We also calculate
anharmonic corrections. The first cumulant is obtained from the net thermal
expansion  yielding 0.0217~\AA\ for fcc Cu at 300K, in accord with the
experimental value $0.0205\pm0.0009$ \AA ~\cite{fornasini04}. The third cumulant
is then estimated from relations among the cumulants \cite{frenkel93}. Further
details are given elsewhere \cite{vila06}.

In summary, we have developed efficient, {\it ab initio} approaches for
calculations of the key many-body damping factors in core-level x-ray and
electron spectroscopies.  Our calculations of inelastic losses are based on a
many-pole representation of the dielectric function and an extension of our RSGF
code that includes corrections to the independent-particle approximation.  This
approach yields significantly improved self-energies compared to single-pole
models, as well as quantitative IMFPs and estimates of losses due to
multi-electron excitations.  We have also developed an efficient approach for
calculations of phonon spectra and MS DW factors. All of these many-body
effects are important in applications to optical constants over a broad
spectrum, and yield improved amplitudes and phases from the UV to x-ray energies
\cite{prangetables05}. Since our approach includes solid state effects (e.g.,
edge shifts, fine structure, and temperature dependent DW factors) and is
applicable to general aperiodic materials, these results complement and can
potentially replace empirical tables
\cite{hagemann75,palik85,henke93,chantler95,elam02} or atomic models
\cite{liberman65} for many applications. Extensions, e.g, to spectra at finite
momentum transfer and to the visible regime with improved potentials are in
progress.

\begin{acknowledgments}
We wish to thank
G. Bertsch,
C. Chantler,
X. Gonze,
H. Krappe,
H. Lawler,
Z. Levine, 
C. Powell,
G. Rivas,
H. Rossner,
E. Shirley,
A. Soininen,
and Y. Takimoto
for comments and suggestions. This work is supported in part by the DOE Grant
DE-FG03-97ER45623 (JJR, MPP) and DE-FG02-04ER1599 (FDV), NIH NCRR BTP Grant
RR-01209 (JJK), and NIST Grant 70 NAMB 2H003 (APS) and was facilitated by the
DOE Computational Materials Science Network.
\end {acknowledgments}


\begin{thebibliography}{38}
\expandafter\ifx\csname natexlab\endcsname\relax\def\natexlab#1{#1}\fi
\expandafter\ifx\csname bibnamefont\endcsname\relax
  \def\bibnamefont#1{#1}\fi
\expandafter\ifx\csname bibfnamefont\endcsname\relax
  \def\bibfnamefont#1{#1}\fi
\expandafter\ifx\csname citenamefont\endcsname\relax
  \def\citenamefont#1{#1}\fi
\expandafter\ifx\csname url\endcsname\relax
  \def\url#1{\texttt{#1}}\fi
\expandafter\ifx\csname urlprefix\endcsname\relax\def\urlprefix{URL }\fi
\providecommand{\bibinfo}[2]{#2}
\providecommand{\eprint}[2][]{\url{#2}}

\bibitem[{\citenamefont{Onida et~al.}(2002)\citenamefont{Onida, Reining, and
  Rubio}}]{onida02}
\bibinfo{author}{\bibfnamefont{G.}~\bibnamefont{Onida}},
  \bibinfo{author}{\bibfnamefont{L.}~\bibnamefont{Reining}}, \bibnamefont{and}
  \bibinfo{author}{\bibfnamefont{A.}~\bibnamefont{Rubio}},
  \bibinfo{journal}{Rev. Mod. Phys.} \textbf{\bibinfo{volume}{74}},
  \bibinfo{pages}{601} (\bibinfo{year}{2002}).

\bibitem[{\citenamefont{Rehr and Albers}(2000)}]{rehr00}
\bibinfo{author}{\bibfnamefont{J.~J.} \bibnamefont{Rehr}} \bibnamefont{and}
  \bibinfo{author}{\bibfnamefont{R.~C.} \bibnamefont{Albers}},
  \bibinfo{journal}{Rev. Mod. Phys.} \textbf{\bibinfo{volume}{72}},
  \bibinfo{pages}{621} (\bibinfo{year}{2000}).

\bibitem[{\citenamefont{Soininen and Shirley}(2001)}]{shirley01}
\bibinfo{author}{\bibfnamefont{J.~A.} \bibnamefont{Soininen}} \bibnamefont{and}
  \bibinfo{author}{\bibfnamefont{E.~L.} \bibnamefont{Shirley}},
  \bibinfo{journal}{Phys. Rev. B} \textbf{\bibinfo{volume}{64}},
  \bibinfo{pages}{165112} (\bibinfo{year}{2001}).

\bibitem[{\citenamefont{Ankudinov et~al.}(2005)\citenamefont{Ankudinov,
  Takimoto, and Rehr}}]{ankudinov05}
\bibinfo{author}{\bibfnamefont{A.~L.} \bibnamefont{Ankudinov}},
  \bibinfo{author}{\bibfnamefont{Y.}~\bibnamefont{Takimoto}}, \bibnamefont{and}
  \bibinfo{author}{\bibfnamefont{J.~J.} \bibnamefont{Rehr}},
  \bibinfo{journal}{Phys. Rev. B} \textbf{\bibinfo{volume}{71}},
  \bibinfo{pages}{165110} (\bibinfo{year}{2005}).

\bibitem[{\citenamefont{Powell and Jablonski}(1999)}]{powell99}
\bibinfo{author}{\bibfnamefont{C.~J.} \bibnamefont{Powell}} \bibnamefont{and}
  \bibinfo{author}{\bibfnamefont{A.}~\bibnamefont{Jablonski}},
  \bibinfo{journal}{J. Phys. Chem. Ref. Data} \textbf{\bibinfo{volume}{28}},
  \bibinfo{pages}{19} (\bibinfo{year}{1999}).

\bibitem[{\citenamefont{Fernandez-Varea
  et~al.}(2005)\citenamefont{Fernandez-Varea, Salvat, Dingfelder, and
  Liljequist}}]{dingfelder05}
\bibinfo{author}{\bibfnamefont{J.~M.} \bibnamefont{Fernandez-Varea}},
  \bibinfo{author}{\bibfnamefont{F.}~\bibnamefont{Salvat}},
  \bibinfo{author}{\bibfnamefont{M.}~\bibnamefont{Dingfelder}},
  \bibnamefont{and}
  \bibinfo{author}{\bibfnamefont{D.}~\bibnamefont{Liljequist}},
  \bibinfo{journal}{Nucl. Instr. and Meth. B} \textbf{\bibinfo{volume}{229}},
  \bibinfo{pages}{187} (\bibinfo{year}{2005}).

\bibitem[{\citenamefont{von~der Linden and Horsch}(1988)}]{horsch88}
\bibinfo{author}{\bibfnamefont{W.}~\bibnamefont{von~der Linden}}
  \bibnamefont{and} \bibinfo{author}{\bibfnamefont{P.}~\bibnamefont{Horsch}},
  \bibinfo{journal}{Phys. Rev. B} \textbf{\bibinfo{volume}{37}},
  \bibinfo{pages}{8351} (\bibinfo{year}{1988}).

\bibitem[{\citenamefont{Krasovskii et~al.}(2002)\citenamefont{Krasovskii,
  Schattke, Strocov, and Claessen}}]{schattke02}
\bibinfo{author}{\bibfnamefont{E.~E.} \bibnamefont{Krasovskii}},
  \bibinfo{author}{\bibfnamefont{W.}~\bibnamefont{Schattke}},
  \bibinfo{author}{\bibfnamefont{V.~N.} \bibnamefont{Strocov}},
  \bibnamefont{and} \bibinfo{author}{\bibfnamefont{R.}~\bibnamefont{Claessen}},
  \bibinfo{journal}{Phys. Rev. B} \textbf{\bibinfo{volume}{66}},
  \bibinfo{pages}{235403} (\bibinfo{year}{2002}).

\bibitem[{\citenamefont{Zangwill and Soven}(1980)}]{zangwill80}
\bibinfo{author}{\bibfnamefont{A.}~\bibnamefont{Zangwill}} \bibnamefont{and}
  \bibinfo{author}{\bibfnamefont{P.}~\bibnamefont{Soven}},
  \bibinfo{journal}{Phys. Rev. A} \textbf{\bibinfo{volume}{21}},
  \bibinfo{pages}{1561} (\bibinfo{year}{1980}).

\bibitem[{\citenamefont{Hagemann et~al.}(1975)\citenamefont{Hagemann, Gudat,
  and Kunz}}]{hagemann75}
\bibinfo{author}{\bibfnamefont{H.~J.} \bibnamefont{Hagemann}},
  \bibinfo{author}{\bibfnamefont{W.}~\bibnamefont{Gudat}}, \bibnamefont{and}
  \bibinfo{author}{\bibfnamefont{C.}~\bibnamefont{Kunz}}, \bibinfo{journal}{J.
  Opt. Soc. Am.} \textbf{\bibinfo{volume}{65}}, \bibinfo{pages}{742}
  (\bibinfo{year}{1975}).

\bibitem[{\citenamefont{Henke et~al.}(1993)\citenamefont{Henke, Gullikson, and
  Davis}}]{henke93}
\bibinfo{author}{\bibfnamefont{B.~L.} \bibnamefont{Henke}},
  \bibinfo{author}{\bibfnamefont{E.~M.} \bibnamefont{Gullikson}},
  \bibnamefont{and} \bibinfo{author}{\bibfnamefont{J.~C.} \bibnamefont{Davis}},
  \bibinfo{journal}{At. Data Nucl. Data Tables} \textbf{\bibinfo{volume}{54}},
  \bibinfo{pages}{181} (\bibinfo{year}{1993}).

\bibitem[{\citenamefont{Prange et~al.}()\citenamefont{Prange, Rivas, Rehr, and
  Ankudinov}}]{prangeetal06}
\bibinfo{author}{\bibfnamefont{M.~P.} \bibnamefont{Prange}},
  \bibinfo{author}{\bibfnamefont{G.}~\bibnamefont{Rivas}},
  \bibinfo{author}{\bibfnamefont{J.~J.} \bibnamefont{Rehr}}, \bibnamefont{and}
  \bibinfo{author}{\bibfnamefont{A.~L.} \bibnamefont{Ankudinov}},
  \bibinfo{note}{unpublished.}

\bibitem[{\citenamefont{Prange et~al.}(2005)\citenamefont{Prange, Rivas, and
  Rehr}}]{prangetables05}
\bibinfo{author}{\bibfnamefont{M.~P.} \bibnamefont{Prange}},
  \bibinfo{author}{\bibfnamefont{G.}~\bibnamefont{Rivas}}, \bibnamefont{and}
  \bibinfo{author}{\bibfnamefont{J.~J.} \bibnamefont{Rehr}},
  \emph{\bibinfo{title}{Tables of Optical Constants}}
  (\bibinfo{publisher}{WWW},
  \bibinfo{address}{\url{http://leonardo.phys.washington.edu/feff/opcons/}},
  \bibinfo{year}{2005}).

\bibitem[{\citenamefont{Rehr et~al.}(2005)\citenamefont{Rehr, Soininen, and
  Shirley}}]{rehr05}
\bibinfo{author}{\bibfnamefont{J.~J.} \bibnamefont{Rehr}},
  \bibinfo{author}{\bibfnamefont{J.~A.} \bibnamefont{Soininen}},
  \bibnamefont{and} \bibinfo{author}{\bibfnamefont{E.~L.}
  \bibnamefont{Shirley}}, \bibinfo{journal}{Physica Scripta}
  \textbf{\bibinfo{volume}{T115}}, \bibinfo{pages}{207} (\bibinfo{year}{2005}).

\bibitem[{\citenamefont{Hedin and Lundqvist}(1969)}]{hedin69}
\bibinfo{author}{\bibfnamefont{L.}~\bibnamefont{Hedin}} \bibnamefont{and}
  \bibinfo{author}{\bibfnamefont{S.}~\bibnamefont{Lundqvist}},
  \bibinfo{journal}{Solid State Phys.} \textbf{\bibinfo{volume}{23}},
  \bibinfo{pages}{1} (\bibinfo{year}{1969}).

\bibitem[{\citenamefont{Lundqvist}(1967)}]{lundqvist67}
\bibinfo{author}{\bibfnamefont{B.}~\bibnamefont{Lundqvist}},
  \bibinfo{journal}{Phys. Kondens. Materie} \textbf{\bibinfo{volume}{6}},
  \bibinfo{pages}{193} (\bibinfo{year}{1967}).

\bibitem[{\citenamefont{Soininen et~al.}(2005)\citenamefont{Soininen,
  Ankudinov, and Rehr}}]{soininen05}
\bibinfo{author}{\bibfnamefont{J.~A.} \bibnamefont{Soininen}},
  \bibinfo{author}{\bibfnamefont{A.~L.} \bibnamefont{Ankudinov}},
  \bibnamefont{and} \bibinfo{author}{\bibfnamefont{J.~J.} \bibnamefont{Rehr}},
  \bibinfo{journal}{Phys. Rev. B} \textbf{\bibinfo{volume}{72}},
  \bibinfo{pages}{045136} (\bibinfo{year}{2005}).

\bibitem[{\citenamefont{Soininen et~al.}(2003)\citenamefont{Soininen, Rehr, and
  Shirley}}]{soininen03}
\bibinfo{author}{\bibfnamefont{J.~A.} \bibnamefont{Soininen}},
  \bibinfo{author}{\bibfnamefont{J.~J.} \bibnamefont{Rehr}}, \bibnamefont{and}
  \bibinfo{author}{\bibfnamefont{E.~L.} \bibnamefont{Shirley}},
  \bibinfo{journal}{J. Phys.: Condens. Matter} \textbf{\bibinfo{volume}{15}},
  \bibinfo{pages}{2572} (\bibinfo{year}{2003}).

\bibitem[{\citenamefont{Campillo et~al.}(1999)\citenamefont{Campillo, Pitarke,
  Rubio, Zarate, and Echenique}}]{rubio99}
\bibinfo{author}{\bibfnamefont{I.}~\bibnamefont{Campillo}},
  \bibinfo{author}{\bibfnamefont{J.~M.} \bibnamefont{Pitarke}},
  \bibinfo{author}{\bibfnamefont{A.}~\bibnamefont{Rubio}},
  \bibinfo{author}{\bibfnamefont{E.}~\bibnamefont{Zarate}}, \bibnamefont{and}
  \bibinfo{author}{\bibfnamefont{P.~M.} \bibnamefont{Echenique}},
  \bibinfo{journal}{Phys. Rev. Lett.} \textbf{\bibinfo{volume}{83}},
  \bibinfo{pages}{2230} (\bibinfo{year}{1999}).

\bibitem[{\citenamefont{Kwei et~al.}(1993)\citenamefont{Kwei, Chen, Tung, and
  Wang}}]{kwei93}
\bibinfo{author}{\bibfnamefont{C.~M.} \bibnamefont{Kwei}},
  \bibinfo{author}{\bibfnamefont{Y.~F.} \bibnamefont{Chen}},
  \bibinfo{author}{\bibfnamefont{C.~J.} \bibnamefont{Tung}}, \bibnamefont{and}
  \bibinfo{author}{\bibfnamefont{J.~P.} \bibnamefont{Wang}},
  \bibinfo{journal}{Surf. Sci.} \textbf{\bibinfo{volume}{293}},
  \bibinfo{pages}{202} (\bibinfo{year}{1993}).

\bibitem[{\citenamefont{Kas et~al.}()\citenamefont{Kas, Campbell, Prange, Rehr,
  and Sorini}}]{kasetal06}
\bibinfo{author}{\bibfnamefont{J.~J.} \bibnamefont{Kas}},
  \bibinfo{author}{\bibfnamefont{L.~W.} \bibnamefont{Campbell}},
  \bibinfo{author}{\bibfnamefont{M.~P.} \bibnamefont{Prange}},
  \bibinfo{author}{\bibfnamefont{J.~J.} \bibnamefont{Rehr}}, \bibnamefont{and}
  \bibinfo{author}{\bibfnamefont{A.~P.} \bibnamefont{Sorini}},
  \bibinfo{note}{unpublished.}

\bibitem[{\citenamefont{Campbell et~al.}(2002)\citenamefont{Campbell, Hedin,
  Rehr, and Bardyszewski}}]{campbell02}
\bibinfo{author}{\bibfnamefont{L.~W.} \bibnamefont{Campbell}},
  \bibinfo{author}{\bibfnamefont{L.}~\bibnamefont{Hedin}},
  \bibinfo{author}{\bibfnamefont{J.~J.} \bibnamefont{Rehr}}, \bibnamefont{and}
  \bibinfo{author}{\bibfnamefont{W.}~\bibnamefont{Bardyszewski}},
  \bibinfo{journal}{Phys. Rev. B} \textbf{\bibinfo{volume}{65}},
  \bibinfo{pages}{064107} (\bibinfo{year}{2002}).

\bibitem[{\citenamefont{Newville}(1994)}]{newville94}
\bibinfo{author}{\bibfnamefont{M.}~\bibnamefont{Newville}}, Ph.D. thesis,
  \bibinfo{school}{University of Washington} (\bibinfo{year}{1994}).

\bibitem[{\citenamefont{Poiarkova and Rehr}(2001)}]{poiarkova01}
\bibinfo{author}{\bibfnamefont{A.}~\bibnamefont{Poiarkova}} \bibnamefont{and}
  \bibinfo{author}{\bibfnamefont{J.~J.} \bibnamefont{Rehr}},
  \bibinfo{journal}{J. Synchrotron Radiat.} \textbf{\bibinfo{volume}{8}},
  \bibinfo{pages}{313} (\bibinfo{year}{2001}).

\bibitem[{\citenamefont{Krappe and Rossner}(2002)}]{krappe02}
\bibinfo{author}{\bibfnamefont{H.~J.} \bibnamefont{Krappe}} \bibnamefont{and}
  \bibinfo{author}{\bibfnamefont{H.~H.} \bibnamefont{Rossner}},
  \bibinfo{journal}{Phys. Rev. B} \textbf{\bibinfo{volume}{66}},
  \bibinfo{pages}{184303} (\bibinfo{year}{2002}).

\bibitem[{\citenamefont{Vila et~al.}()\citenamefont{Vila, Krappe, Rossner, and
  Rehr}}]{vila06}
\bibinfo{author}{\bibfnamefont{F.~D.} \bibnamefont{Vila}},
  \bibinfo{author}{\bibfnamefont{H.~J.} \bibnamefont{Krappe}},
  \bibinfo{author}{\bibfnamefont{H.~H.} \bibnamefont{Rossner}},
  \bibnamefont{and} \bibinfo{author}{\bibfnamefont{J.~J.} \bibnamefont{Rehr}},
  \bibinfo{note}{unpublished.}

\bibitem[{\citenamefont{Gonze et~al.}(2002)\citenamefont{Gonze, Beuken,
  Caracas, Detraux, Fuchs, Rignanese, Sindic, Verstraete, Zerah, Jollet
  et~al.}}]{gonze02}
\bibinfo{author}{\bibfnamefont{X.}~\bibnamefont{Gonze}},
  \bibinfo{author}{\bibfnamefont{J.-M.} \bibnamefont{Beuken}},
  \bibinfo{author}{\bibfnamefont{R.}~\bibnamefont{Caracas}},
  \bibinfo{author}{\bibfnamefont{F.}~\bibnamefont{Detraux}},
  \bibinfo{author}{\bibfnamefont{M.}~\bibnamefont{Fuchs}},
  \bibinfo{author}{\bibfnamefont{G.-M.} \bibnamefont{Rignanese}},
  \bibinfo{author}{\bibfnamefont{L.}~\bibnamefont{Sindic}},
  \bibinfo{author}{\bibfnamefont{M.}~\bibnamefont{Verstraete}},
  \bibinfo{author}{\bibfnamefont{G.}~\bibnamefont{Zerah}},
  \bibinfo{author}{\bibfnamefont{F.}~\bibnamefont{Jollet}},
  \bibnamefont{et~al.}, \bibinfo{journal}{Computational Materials Science}
  \textbf{\bibinfo{volume}{25}}, \bibinfo{pages}{478} (\bibinfo{year}{2002}).

\bibitem[{\citenamefont{Gonze and Lee}(1997)}]{gonze97}
\bibinfo{author}{\bibfnamefont{X.}~\bibnamefont{Gonze}} \bibnamefont{and}
  \bibinfo{author}{\bibfnamefont{C.}~\bibnamefont{Lee}},
  \bibinfo{journal}{Phys. Rev. B} \textbf{\bibinfo{volume}{55}},
  \bibinfo{pages}{10355} (\bibinfo{year}{1997}).

\bibitem[{\citenamefont{Fuchs and Scheffler}(1999)}]{fuchs99}
\bibinfo{author}{\bibfnamefont{M.}~\bibnamefont{Fuchs}} \bibnamefont{and}
  \bibinfo{author}{\bibfnamefont{M.}~\bibnamefont{Scheffler}},
  \bibinfo{journal}{Comput. Phys. Commun.} \textbf{\bibinfo{volume}{119}},
  \bibinfo{pages}{67} (\bibinfo{year}{1999}).

\bibitem[{\citenamefont{Stern et~al.}(1980)\citenamefont{Stern, Bunker, and
  Heald}}]{stern80}
\bibinfo{author}{\bibfnamefont{E.~A.} \bibnamefont{Stern}},
  \bibinfo{author}{\bibfnamefont{B.~A.} \bibnamefont{Bunker}},
  \bibnamefont{and} \bibinfo{author}{\bibfnamefont{S.~M.} \bibnamefont{Heald}},
  \bibinfo{journal}{Phys. Rev. B} \textbf{\bibinfo{volume}{21}},
  \bibinfo{pages}{5521} (\bibinfo{year}{1980}).

\bibitem[{\citenamefont{Perdew and Wang}(1992)}]{perdew92}
\bibinfo{author}{\bibfnamefont{J.~P.} \bibnamefont{Perdew}} \bibnamefont{and}
  \bibinfo{author}{\bibfnamefont{Y.}~\bibnamefont{Wang}},
  \bibinfo{journal}{Phys. Rev. B} \textbf{\bibinfo{volume}{45}},
  \bibinfo{pages}{13244} (\bibinfo{year}{1992}).

\bibitem[{\citenamefont{Perdew et~al.}(1996)\citenamefont{Perdew, Burke, and
  Ernzerhof}}]{perdew96}
\bibinfo{author}{\bibfnamefont{J.~P.} \bibnamefont{Perdew}},
  \bibinfo{author}{\bibfnamefont{K.}~\bibnamefont{Burke}}, \bibnamefont{and}
  \bibinfo{author}{\bibfnamefont{M.}~\bibnamefont{Ernzerhof}},
  \bibinfo{journal}{Phys. Rev. Lett.} \textbf{\bibinfo{volume}{77}},
  \bibinfo{pages}{3865} (\bibinfo{year}{1996}).

\bibitem[{\citenamefont{Fornasini et~al.}(2004)\citenamefont{Fornasini,
  a~Beccara, Dalba, Grisenti, Sanson, Vaccari, and Rocca}}]{fornasini04}
\bibinfo{author}{\bibfnamefont{P.}~\bibnamefont{Fornasini}},
  \bibinfo{author}{\bibfnamefont{S.}~\bibnamefont{a~Beccara}},
  \bibinfo{author}{\bibfnamefont{G.}~\bibnamefont{Dalba}},
  \bibinfo{author}{\bibfnamefont{R.}~\bibnamefont{Grisenti}},
  \bibinfo{author}{\bibfnamefont{A.}~\bibnamefont{Sanson}},
  \bibinfo{author}{\bibfnamefont{M.}~\bibnamefont{Vaccari}}, \bibnamefont{and}
  \bibinfo{author}{\bibfnamefont{F.}~\bibnamefont{Rocca}},
  \bibinfo{journal}{Phys. Rev. B} \textbf{\bibinfo{volume}{70}},
  \bibinfo{pages}{174301} (\bibinfo{year}{2004}).

\bibitem[{\citenamefont{Frenkel and Rehr}(1993)}]{frenkel93}
\bibinfo{author}{\bibfnamefont{A.~I.} \bibnamefont{Frenkel}} \bibnamefont{and}
  \bibinfo{author}{\bibfnamefont{J.~J.} \bibnamefont{Rehr}},
  \bibinfo{journal}{Phys. Rev. B} \textbf{\bibinfo{volume}{48}},
  \bibinfo{pages}{585} (\bibinfo{year}{1993}).

\bibitem[{\citenamefont{Palik}(1985)}]{palik85}
\bibinfo{editor}{\bibfnamefont{E.~D.} \bibnamefont{Palik}}, ed.,
  \emph{\bibinfo{title}{Handbook of Optical Constants of Solids}}
  (\bibinfo{publisher}{Academic Press, Orlando}, \bibinfo{year}{1985}).

\bibitem[{\citenamefont{Chantler}(1995)}]{chantler95}
\bibinfo{author}{\bibfnamefont{C.~T.} \bibnamefont{Chantler}},
  \bibinfo{journal}{J. Phys. Chem. Ref. Data} \textbf{\bibinfo{volume}{24}},
  \bibinfo{pages}{71} (\bibinfo{year}{1995}).

\bibitem[{\citenamefont{Elam et~al.}(2002)\citenamefont{Elam, Ravel, and
  Sieber}}]{elam02}
\bibinfo{author}{\bibfnamefont{W.~T.} \bibnamefont{Elam}},
  \bibinfo{author}{\bibfnamefont{B.~D.} \bibnamefont{Ravel}}, \bibnamefont{and}
  \bibinfo{author}{\bibfnamefont{J.~R.} \bibnamefont{Sieber}},
  \bibinfo{journal}{Rad. Phys. Chem.} \textbf{\bibinfo{volume}{63}},
  \bibinfo{pages}{121} (\bibinfo{year}{2002}).

\bibitem[{\citenamefont{Liberman et~al.}(1965)\citenamefont{Liberman, Waber,
  and Cromer}}]{liberman65}
\bibinfo{author}{\bibfnamefont{D.}~\bibnamefont{Liberman}},
  \bibinfo{author}{\bibfnamefont{J.~T.} \bibnamefont{Waber}}, \bibnamefont{and}
  \bibinfo{author}{\bibfnamefont{D.~T.} \bibnamefont{Cromer}},
  \bibinfo{journal}{Phys. Rev.} \textbf{\bibinfo{volume}{137}},
  \bibinfo{pages}{A27} (\bibinfo{year}{1965}).

\end{thebibliography}

\end{document}